\documentclass[reqno,11pt]{amsart}
\usepackage{hyperref}
\usepackage{graphicx}
\usepackage{slashed}
\usepackage{amscd}
\usepackage{tensor}
\usepackage{amssymb}
\usepackage{esint}
\usepackage{enumitem}
\usepackage{accents}
\usepackage[dvipsnames]{xcolor}
\usepackage[utf8]{inputenc}
\usepackage[off]{auto-pst-pdf}
\usepackage{pst-grad}
\usepackage{pst-plot}
\usepackage[mathscr]{eucal}
\usepackage{soul}	

\usepackage{tagging}
\droptag{NotReady}

\newcommand\Johannes[1]{}
\renewcommand\Johannes[1]{\marginpar {\flushleft\sffamily\footnotesize {\textcolor{CornflowerBlue}{Johannes}}: #1}}

\allowdisplaybreaks[1]

\newtheorem{Def}{Definition}
\newtheorem{Thm}{Theorem}

\newtheorem{Example}[Def]{Example}

\newcommand{\beq}{\begin{equation}}
\newcommand{\eeq}{\end{equation}}
\newcommand{\Proof}{\begin{proof}}
\newcommand{\QED}{\end{proof} \noindent}
\newcommand{\QEDrem}{\ \hfill $\Diamond$}

\newcommand{\M}{{\mathbb{M}}}

\renewcommand{\O}{{\mathscr{O}}}

\DeclareFontFamily{OT1}{rsfso}{}
\DeclareFontShape{OT1}{rsfso}{m}{n}{ <-7> rsfso5 <7-10> rsfso7 <10-> rsfso10}{}
\DeclareMathAlphabet{\mycal}{OT1}{rsfso}{m}{n}

\newcommand{\bei}{\begin{itemize}[label=$\circ$,itemsep=.5em,leftmargin=*]}
\newcommand{\beii}{\begin{itemize}[label=-,itemsep=.5em,topsep=.5em,leftmargin=*]}
\newcommand{\eni}{\end{itemize}}
\newcommand{\enii}{\end{itemize}}

\renewcommand{\P}{\mathscr P}
\newcommand{\PP}{\mathbb P}
\newcommand{\OO}{\mathbb O}

\definecolor{darkblue}{RGB}{0,91,163}

\usepackage[onehalfspacing]{setspace}
\usepackage[numbers]{natbib}

\title[The Closure of the Physical is Unscientific]{The Closure of the Physical is Unscientific}
\author[J. Kleiner]{Johannes Kleiner}
\author[S. Hartmann]{Stephan Hartmann}

\begin{document}

\noindent

\noindent
\centerline{\LARGE \huge
The Closure of the Physical,}\\[1em]
\centerline{\LARGE \huge
Consciousness and Scientific Practice}\\[1em]

\vspace*{.8cm}

\centerline{\Large Johannes Kleiner$^{1,2}$ and Stephan Hartmann$^{1,3}$}
\vspace*{.3cm}
\centerline{$^1$Munich Center for Mathematical Philosophy}
\centerline{$^2$Munich Graduate School of Systemic Neurosciences}
\centerline{$^3$Munich Center for NeuroSciences -- Brain \& Mind}

\vspace*{3em}
\begin{quote}\small
\textsc{Abstract}
We analyse the implications of the closure of the physical for experiments in the scientific study of consciousness when all the details are considered, especially how measurement results relate to physical events. It turns out that the closure of the physical has surprising implications that conflict with scientific practice. These implications point to a fundamental flaw in the paradigm underlying many experiments conducted to date and pose a challenge to any research programme that aims to ground a physical functionalist or identity-based understanding of consciousness on empirical observations.
\end{quote}

\vspace*{2em}

\section{Introduction}

The closure of the physical is a central assumption in the philosophy of mind and in the scientific study of consciousness~\cite{kim1996philosophy,papineau2009causal}. It underlies both functionalist and identity theories of consciousness and is a central component of many, if not all, neuroscientific models of consciousness. However, we will show below that the closure of the physical is untenable in a scientific context because it implies that no experiment can actually distinguish between two theories of consciousness that obey this assumption. It is therefore incompatible with scientific practice and hence \emph{unscientific}.

The central idea of our argument is the observation that in any scientific experiment the measurement results must be stored or transmitted before analysis, and we show that this means that the stored data are determined by the physical properties of a storage device or a transmission channel. In conjunction with the closure of the physical, this means that the stored data are independent of which theory of consciousness is true.

It has already been pointed out that the closure of the physical is a problematic assumption in a scientific context. \cite{pauen2000painless} and~\cite{pauen2006feeling}, for example, make this point with respect to property dualism and qualia epiphenomenalism. Our proof presented below covers the general case. It shows independently of any other metaphysical premises that one of the central assumptions in the empirical study of consciousness is flawed. This calls into question the theoretical basis of a large number of experiments conducted to date and shows that the hope of basing a physical functionalist or identity-based understanding of consciousness on empirical observations is null and void.

The remainder of this paper is organized as follows. Section~\ref{sec:theories} elaborates which theories of consciousness our argument addresses and defines an epistemic version of the closure of the physical. Section~\ref{sec:experiments} identifies a necessary condition for theories of consciousness to be distinguished by empirical data. Sections~\ref{sec:data} and~\ref{sec:measurement-results} discuss the role of empirical data in the scientific study of consciousness and why they supervene on physical events. Section~\ref{sec:argument} is devoted to the proof of our main claim, and Section~\ref{sec:cc} shows that the causal closure of the physical, as usually defined ontologically, implies our definition, which ensures that our result holds in full generality. Finally, Section~\ref{sec:final} contains some concluding remarks.

\section{Theories of consciousness}\label{sec:theories}

We use the term \emph{theories of consciousness} to refer to the theories that are tested, compared, or derived in experiments in the scientific study of consciousness, regardless of what metaphysical status of consciousness they presuppose. This includes, for example, Integrated Information Theory~\cite{oizumi2014phenomenology}, Global Neuronal Workspace Theory~\cite{mashour2020conscious} or Higher Order Thought Theory~\cite{brown2019understanding}, and in general all scientific theories which adhere to functionalism, identity theory or epiphenomenalism. This also includes illusionist or eliminativist theories
that are subject to experimental testing, even though they do not
grant consciousness an independent ontological status, but merely aim to
explain why someone has the illusion of being conscious~\cite{sprevak2020eliminativism}.

Our results rely on two general facts about theories of consciousness. The first is that theories of consciousness relate to physical events, where \emph{physical events} are the kinds of events that are the subject of natural sciences such as biology, chemistry, neuroscience, and physics. Some theories modify the description of physical events provided by natural science, for example, by postulating changes in the temporal evolution of physical states, as recently in~\cite{chalmers2021consciousness}, others simply adopt whatever natural science says about physical events without any modification.

The causal closure of the physical is the assumption that for every physical effect, there is a sufficient physical cause. Its key epistemic repercussion (cf. Section~\ref{sec:cc}) is that theories of consciousness must not amend whatever it is that the physical sciences say or imply about physical events. We call this epistemic assumption \emph{closure of the physical}:  A theory of consciousness obeys the \emph{closure of the physical} if and only if it does not posit any changes to the physical events explained, predicted or otherwise determined by natural science.

This premise can be expressed concisely in formal terms. To this end, we introduce two sets\footnote{Note that we do not distinguish between classes and sets in this paper.} of event-descriptions. First, for any theory of consciousness $T$, we denote by $\P_T$ the physical events which $T$ is committed to, for example the firing of some neurons or the instantiation of some functional property.  Every element in $\P_T$ is a description of an event that occurs, according to $T$, in the actual world. The description specifies the event and may include properties or relational information about the event. What exactly a description contains and in which language it is formulated is not of importance here.

Second, we denote by $\P_P$ the physical events which natural science explains, predicts or determines. Whatever it is that natural science says or implies about the physical events in the actual world is part of the class $\P_P$. Each element is in turn a description of an event, including its properties and relations, and we allow that the description is either deterministic or indeterministic.%
\footnote{In terms of a fundamental physical theory, $\P_P$ may be thought of as comprising all events which are part of those dynamically possible trajectories that occur in the actual world.}

Since scientific theories are complex, $\P_P$ may not be known or even knowable. And as science progresses, $\P_P$ is likely to change over time. For this reason, in what follows,~$\P_P$ functions like a variable. It is not important what value this variable actually takes, but only what relationship a theory of consciousness has to this variable.

A theory of consciousness obeys the closure of the physical only if it does not postulate any changes to the class $\P_P$. Thus, it does not replace, change, or add to the description of physical events explained, predicted, or otherwise determined by natural science. This means that for every physical event in $\P_T$ to which a theory of consciousness is committed, there is an element of $\P_P$ that provides a description of that event in one of the languages of a natural science. The descriptions in the two sets may differ in language, but not in content.

In formal terms, this means that there is an \emph{embedding} of $\P_T$ into $\P_P$, i.e. an injective (one-to-one) function~$\iota$ of the form
\begin{equation}\label{eq:iota}
\iota: \P_T \longrightarrow \P_P \:,
\end{equation}
which specifies for every physical event and description that the theory of consciousness is committed to the corresponding event and description explained,  predicted, or determined by natural science.
The existence of this function is the concise meaning of the closure of the physical introduced above: A theory of consciousness $T$ obeys the \emph{closure of the physical} if and only if there exists a function $\iota$ as in~\eqref{eq:iota}. We will show in Section~\ref{sec:cc} that the usual reading of the causal closure of the physical implies just that.%
\footnote{
The closure of the physical so conceived could also be defined in terms of variables and other concepts used in scientific theories, such that a theory of consciousness obeys the closure of the physical if and only if it makes no change to the concepts that natural scientific theories employ to predict and explain physical events, or which otherwise determine physical events. While this formulation would capture the more familiar assumption that ``physical laws already
form a closed system''~\cite[p.\,127]{chalmers1996conscious}, it introduces another level of abstraction (concepts used in scientific theories) that is avoided when formulated in terms of events.
}
\section{Experiments}\label{sec:experiments}

In the scientific study of consciousness, experiments are conducted to falsify, confirm, or distinguish between competing theories of consciousness.
The most important component of any experiment is measurement, i.e., laboratory operations that produce a set of data which constitutes the result of the measurement.

The second general fact on which our argument is based is that scientific theories of consciousness have something to say about possible measurement results. We assume that any theory allows one to derive, for some experiments and under appropriate auxiliary assumptions, a class of data sets which, according to the theory, may occur as the result of the experiment. This requirement singles out \emph{scientific} theories as those to which our argument applies.%
\footnote{In particular, if we assume that experiments are required to distinguish between competing theories of consciousness, we assume that consciousness cannot be deduced from the physical or, if it can, that experiments are required to figure out how because the deduction fails in practice due to complexity and/or too little knowledge.}

We use the symbol $\M$ to represent an experiment, 
and furthermore introduce the symbol $\O_{\M}$ to denote all data sets which could result from this experiment according to some assumption or theory. So $\O_{\M}$ denotes the possible measurement results of $\M$ in some context. If an experiment $\M$ only made measurements on one system and everything were deterministic, then there would only be one data set in $\O_{\M}$. But experiments usually consider many systems and things are not deterministic, which is why we have a whole class of data sets that can occur in $\M$.%
\footnote{
For now, $\M$ can be thought of as an experiment actually conducted in the actual world to distinguish between theories of consciousness, although logical possibilities will come into play in Section~\ref{sec:data}.}

Given an experiment $\M$ to which a theory $T$ can be applied, we denote the data sets which can occur in $\M$ according to $T$ by $\O_T$. In experimental practice, $\O_T$ is deduced from~$T$, making use of approximations and auxiliary assumptions, so that it contains the pre- or retrodictions of $T$. But in our case we stick to the precise meaning independently of approximations and auxiliary assumptions. Any result $o \in \O_T$ can occur in experiment $\M$ after $T$, and any $o \not\in \O_T$ cannot occur in $\M$ after $T$. If $o \in \O_T$ occurs, then the probability of $T$ increases (and $T$ is confirmed), and if $o \not\in \O_T$ occurs, then the probability of $T$ decreases (and $T$ is disconfirmed). In a Popperian framework, the occurrence of $o \in \O_T$ provides a corraboration of $T$ and the occurrence of $o \not\in \O_T$ amounts to a falsification of $T$.

What matters for our purposes is that if two theories provide the exact same information about which results may or may not occur in an experiment, then these theories cannot be distinguished in that experiment. Theories for which this is the case are empirically indistinguishable. Put concisely in terms of the notation we have just introduced, two theories $T$ and $T'$ are \emph{empirically indistinguishable} if there is no single experiment $\M$ such that $\O_T \neq \O_{T'}$ in $\M$. So if two theories are to be empirically distinguishable, they cannot yield exactly the same class of possible measurement results for each experiment. There must be at least one experiment in which $\O_T \neq \O_{T'}$, so that in this experiment there is at least a chance that a result $o$ occurs which lies in one but not in both classes and is thus consistent with one but not with both theories.\footnote{Note that empirical indistinguishability is weaker than empirical equivalence, as defined, for example, in~\cite{weatherall2019part1} and~\cite{weatherall2019part2}. Two theories are empirically indistinguishable if they make exactly the same testable statements about experiments to which they are both applicable. Empirical equivalence also requires that the two theories apply to exactly the same experiments.}

It is natural to expect that a large number of experiments will not be able to distinguish between two arbitrary theories, since experiments are usually designed with specific theories in mind. Empirical indistinguishability holds only if for two theories there is no experiment at all that can distinguish between them.

If an assumption implies that this is in fact true of \emph{all} theories obeying this assumption, and if there are two or more competing theories which do so, this is obviously problematic. In case such an assumption is implied, all experiments that seek to distinguish between theories become meaningless, and all subsequent differences between theories obeying that assumption untestable. This is incompatible with any empirically based scientific practice, so we take this a sufficient condition to call such an assumption unscientific. Thus, if an assumption implies that any two different theories obeying that assumption are empirically indistinguishable, we conclude that the assumption is \emph{unscientific}.

We emphasize that this condition is a decidedly weak sufficient condition for a particular assumption not to be scientific. We have by no means proposed a new solution to the notorious demarcation problem. Moreover, the condition is independent of the choice of the preferred account of theory testing. An assumption that is unscientific in this sense undermines any empirical scientific progress in the field in question.

Experiments in the scientific study of consciousness usually use two different types of measurements~\cite{chalmers2004can}. First, they make use of what are called \emph{third-person measurements} which employ standard scientific methods. Typical examples are EEG or fMRI recordings. Second, they use what might be called \emph{first-person} or \emph{consciousness-inferring} measurements. This class of measurements has been characterized as using the subject's access to his or her own conscious experience in some way, such as via verbal reports or pressing of a button~\cite{metzinger1995problem}. More recently, the term \emph{subjective measures of consciousness} has come to refer to these types of measures~\cite{irvine2013measures}, in contrast to \emph{objective measures} and \emph{no-report paradigms}~\cite{tsuchiya2015no}, which infer a subject's state of consciousness indirectly, e.g., by evaluating forced choice tasks~\cite{del2007brain} or behavioral data such as 
optokinetic nystagmus and the pupillary reflex~\cite{frassle2014binocular}.

What exactly the difference is between measurements in the first and third person is not important for our purposes. 
The only important thing is that both types of measurements produce results that need to be analyzed, interpreted or transformed. To do this, they must be stored on a data repository. This fact has implications that we analyze below.%
\footnote{We emphasize that this also holds true for ``measuring'' consciousness by introspection. Because science is an intersubjective endeavor, whatever is accessed by introspection in any experiment that aims to distinguish among competing theories of consciousness has to be stored or transmitted in order to be shared with other scientists.
Nothing hinges on how precisely one flashes out what is special about consciousness and its measurement. What matters below is only that measurement results need to be stored or transmitted and that different theories of consciousness may be formulated which are compatible with the same set of physical events. The closure of the physical enforces the latter.
}

\section{Data}\label{sec:data}

We have minimally characterized measurements as laboratory operations that provide a data set that is designated as the result of the experiment. But what does it mean that this data set must be stored on some device? To address this question, let's take a hard disk as an example. A hard disk stores data by magnetizing a thin film of ferromagnetic material that forms the surface of the hard disk platter. The film is made up of many tiny, sequentially aligned magnetic regions, each of which has a magnetization vector that can point in one of two directions. When data is stored on the disk, the head of the drive arm moves over these areas and changes the magnetization vector by applying electric fields. When reading data from the disk, the actuator arm uses weaker electric fields to sense the magnetization vectors of the areas.

The data stored on the disk is the distribution of magnetization vectors across the magnetic areas in terms of the order of the areas. Two copies of the same disk cannot differ in the data stored on it without differing in at least some magnetization vectors. The data is \emph{determined} by the magnetization vectors.

The crucial thing about the magnetization vectors that determine the data stored on a hard disk is that they are not just properties of the device, but actually \emph{physical properties} of the device, the kind of properties that are the subject of natural science, in this case electromagnetism. Electromagnetism explains their causal properties, such as how the magnetization vector responds to electric fields, and also their dynamic properties, such as how magnetization vectors change over time without interactions.

Accordingly, the occurrence of a particular distribution of magnetization vectors over the ferromagnetic film at a particular time is a \emph{physical event}, the kind of event that is the subject of natural science. It follows that the data stored on the hard disk is determined by a physical event: in this case, the distribution of magnetization vectors over the ferromagnetic film. There is no constraint on why or how this physical event occurs, but once the event occurs, the data stored on the hard disk is determined.

This is true not only for hard drives, but for all data storage devices, such as solid-state drives or flash drives, where the relevant semiconductor properties can only be explained using condensed matter theory and quantum mechanics. But even when data is stored on something as simple as a piece of paper or a spoken word, the data supervene on physical events, namely the distribution of ink molecules on the paper material and air pressure fluctuations, which in these cases represent sound waves.

We can again express this fact succinctly in formal terms. Functions in the mathematical sense of the word are defined to capture exactly those cases where something is completely determined by something else. Let us denote by $\PP$ the set or class of all physical events (and descriptions) that can possibly occur in the real world, and by~$\OO_D$ all records that can possibly be stored on a storage device $D$. The notion of possibility at issue here is logical possibility. The physical events explained, predicted, or determined by natural science for the actual world form a subset of $\PP$, the subset $\P_P$ we introduced above. The same is true for the physical events $\P_T$ to which a theory of consciousness is committed. 

The fact that the physical events which occur in the actual world determine the data that is stored on a storage device $D$ can then be represented by a function
\begin{equation}\label{eq:dDP}
d_D: P(\PP) \longrightarrow P(\OO_D) \:,
\end{equation}
where $P(\PP)$ is the set of all subsets of $\PP$, called the power set of $\PP$, and where $P(\OO_D)$ is the power set of $\OO_D$.
The function $d_D$ provides for every logically possible set of physical events $\P \subset \PP$ of the actual world a class of data sets $\O_D \subset \OO_D$ that could be stored on $D$ at a particular time, so it maps element-wise as
\begin{align}\label{eq:dDPmapsto}
d_D: \P \longmapsto \O_D \:.
\end{align}
It selects from all physical events which, according to $\P$, are part of the real world those which are relevant for data storage on the device $D$, e.g. the magnetization vectors in the case of a hard disk. Since $\P$ may contain indeterministic statements, the output of the function may also be indeterministic. For this reason, the output is represented by a class $\O_D$, which may contain more than one record $o$. However, although $\O_D$ is consistent with indeterminism in physical events, it is completely determined by $\P_P$. This is enforced by the fact that $d_D$ is a function. If $D$ is not instantiated in a set $\P$, the function simply returns the empty set.

In order to use this function in the following, we have to consider two conditions. The first condition arises from the fact that the data stored on a device $D$ corresponding to some physical events is independent of the language used to describe those events. Applied to the embedding $\iota$ introduced in~\eqref{eq:iota}, this means that
\begin{equation}\label{eq:language-independence}
d_D\big( \iota(\P_T) \big) = d_D\big( \P_T \big) \:.
\end{equation}
The content of $\iota(\P_T)$ and $\P_T$ is the same, so also the data stored on $D$.

The second condition targets situations where one set of physical events completely contains another, e.g. when the latter is a partial description of the former. A set of physical events $\P_2$ completely contains another set $\P_1$ if all event descriptions of $\P_1$ are also contained in $\P_2$, which means that $\P_2$ describes exactly the same events as $\P_1$. It may add to the description of $\P_1$, but it does not change it in any way. Thus, if $\P_1$ includes all the physical events required to instantiate a data repository $D$, and thus determines the data stored on $D$, it follows that $\P_2$ also includes these events, so that the data that $\P_1$ and $\P_2$ determine to be stored on $D$ are the same. Whenever we have $\P_1 \subset \P_2$ and $D$ is instantiated in $\P_1$, we have
\begin{equation}\label{eq:instantiated}
d_D \big( \P_1 \big) = d_D \big( \P_2 \big) \:.
\end{equation}

\section{Measurement results}\label{sec:measurement-results}

We are now ready to apply this result on data storage to experiments in the scientific study of consciousness. The measurements performed in these experiments tend to be quite complex. They may employ advanced brain imaging techniques such as EEG, ECoG, or fMRI, and require finely tuned equipment and sophisticated analysis to learn about a subject's state of consciousness.

In the case of EEG, ECoG or fMRI recordings, it is relatively clear what the result of such measurements is. It is the data set that the scanner provides after each trial and that is stored in computer memory. In the case of subjective measures, one would normally expect reports or keystrokes to count as results; in the case of objective measures, changes in pupil size and the like. Crucially, however, all of these are physical events. The electrical activity that an EEG electrode measures is as much a physical event as the sound waves that make up a spoken word or the mechanical movements of a button.

Our analysis from the last section allows us to make this point despite the terminological ambiguities about what to count as the result of a measurement. A necessary condition for a record to count as the result of a measurement is that it be stored somewhere. This can be computer memory, but it can also be something simpler like ink on paper or density fluctuations in sound waves. Even data transmission, such as in a cable attached to a button that a person presses, is a form of data storage, albeit of very short duration. So for something to be considered a measurement at all, there must necessarily be a data repository $D$, so that some of the data stored on $D$ \emph{is} the result of the measurement.

However, we have established above that the data stored on a device $D$ is determined by physical events. Since a part of this data represents the measurement result, the measurement results are also determined by physical events. How these physical events come about -- what their causes are -- is not constrained by our analysis. The events can have purely physical causes, physical and non-physical causes, or a priori only non-physical causes. Which of these cases applies and with respect to which notion of causality depends on the theory of consciousness.

As before, let us denote by $\M$ an arbitrary but fixed experiment in the scientific study of consciousness, and let us denote by $D$ the data store or stores necessarily used in $\M$ to store the results of the measurement. We have already introduced the symbol $\O_{\M}$ to denote the data sets that, under certain assumptions or theories, could be the possible outcomes of the experiment $\M$. Our analysis from the previous section then shows that $\O_{\M}$ is also determined by the function $d_D$ introduced in~\eqref{eq:dDP}, namely by restricting $d_D$ to the part of the data stored on $D$ that represents the measurement results. If we denote this restriction by $d_{\M}$ and all data sets that could possibly result from $\M$ by $\OO_{\M}$, we obtain a function
\begin{align}\begin{split}\label{eq:dMmapsto}
d_{\M}: P(\PP) &\longrightarrow P(\OO_{\M})\\
\P &\longmapsto \O_{\M }\:,
\end{split}\end{align}
which maps any set of physical events~$\P$, which could possibly represent the physical events of the actual world, to the measurement results, which in this case would be determined as the result of the experiment $\M$.

The function $d_{\M}$ establishes a connection between what a theory of consciousness~$T$ predicts or postulates about physical events in the real world, on the one hand, and the possible measurement outcomes that can occur according to~$T$, on the other. It selects from the events $\P_T$ that the theory $T$ is committed to those events which determine the data that is stored on $D$. Making use of the symbol $\O_T$ introduced above to denote the possible measurement results that can occur in $\M$ after $T$, this means that
\begin{equation}\label{eq:dPTOT}
    d_{\M}(\P_T) = \O_T \:.
\end{equation}
In this way, we can determine $\O_T$ independently of approximations or auxiliary assumptions.

\section{Why the closure of the physical is unscientific}\label{sec:argument}

By considering that measurement results must be stored and are thereby determined by physical events, we have obtained a novel, additional handle for analyzing experiments in the scientific study of consciousness. In addition to what experimenters derive from a theory $T$ and appropriate auxiliary assumptions, we can now analyze measurement results along the path of what a theory of consciousness says about physical events. This gives rise to the following theorem.

\begin{Thm}\label{thm:closure-unscientific}
The closure of the physical is unscientific.
\end{Thm}

\Proof
Let $T_1$ and $T_2$ denote two theories of consciousness which obey the closure of the physical. This implies that there exist embeddings $\iota_1: \P_{T_1} \longrightarrow \P_P$ and $\iota_2: \P_{T_2} \longrightarrow \P_P$ as in~\eqref{eq:iota}. Let $\M$ denote an experiment to which both $T_1$ and $T_2$ are applicable, and $D$ the data storage device(s) used in that experiment. 
Because of condition~\eqref{eq:language-independence}, we have $d_D( \iota_1(\P_{T_1}) ) = d_D( \P_{T_1} ) $ and $d_D( \iota_2(\P_{T_2}) ) = d_D( \P_{T_2} ) $. 

Both $T_1$ and $T_2$ need to be committed to the existence of physical events which instantiate the data storage device $D$ used in $\M$, for otherwise they would violate the very conditions that make $\M$ possible. Therefore, $D$ is instantiated in both $\P_{T_1}$ and $\P_{T_2}$. Because applying $\iota_1$ resp. $\iota_2$ does not change the content of the described events, it follows that $D$ is also instantiated in $\iota_1(\P_{T_1}) $, resp. $\iota_2(\P_{T_2}) $. 

Because $\iota_1$ is an embedding, we have $\iota_1( \P_{T_1} ) \subset \P_P$. Because $D$ is instantiated in 
$\iota_1( \P_{T_1} )$, Equation~\eqref{eq:instantiated} applies so that we have $d_D( \iota_1( \P_{T_1} ) ) = d_D(\P_P)$. The same applies to $\iota_2$, so that also here, Equation~\eqref{eq:instantiated} implies $d_D( \iota_2( \P_{T_2} ) ) = d_D(\P_P)$. So we in fact have $d_D( \iota_1( \P_{T_1} ) ) = d_D( \iota_2( \P_{T_2} ) )$, which in light of the above implies $d_D( \P_{T_1} )  = d_D( \P_{T_2} ) $. 

We thus find that the data stored on $D$ is exactly the same for both theories. Restriction to $d_{\M}$ introduced in~\eqref{eq:dMmapsto} furthermore implies that $d_{\M}(\P_{T_1}) = d_{\M}(\P_{T_2})$, and because of~\eqref{eq:dPTOT}, this implies that $\O_{T_1} = \O_{T_2}$. So the measurement results of $\M$ are exactly the same according to both $T_1$ and $T_2$. Independently of which predictions one arrives at by making use of auxiliary assumptions, the closure of the physical implies that the data sets which can occur in $\M$ cannot differ.

Since $\M$ was chosen arbitrarily, this conclusion holds for any experiment $\M$, so $T_1$ and $T_2$ are empirically indistinguishable. And because $T_1$ and $T_2$ were arbitrarily chosen among the theories obeying the closure of the physical, we can conclude that all theories obeying the closure of the physical are empirically indistinguishable. It follows that the closure of the physical is an assumption that is unscientific.
\QED

\section{Causal closure of the physical}\label{sec:cc}

The \emph{causal closure of the physical} is the assumption that for every physical effect there is a sufficient physical cause.
This is an ontological assumption; it refers to what is the case in the actual world. In contrast, the assumption we have been working with above -- that a theory of consciousness obeys the \emph{closure of the physical} if and only if it does not postulate changes in physical events explained, predicted, or otherwise determined by natural science -- is epistemic in nature, it depends on the definition, formulation, and content of a theory of consciousness.

The precise meaning of the causal closure of the physical depends heavily on what notion of causality one subsumes, what ontology one grants to causality (if any), and what one allows as relata of the causal relation. Nevertheless, there is a great deal of consensus about what epistemic implications this assumption has.

According to Jaegwon Kim, for example, the causal closure of the physical implies that ``to explain the occurrence of a physical event we never need to go outside of the physical realm''~\cite[p.\,147]{kim1996philosophy}. And Frank Jackson characterizes the causal closure of the physical as the claim that ``the physical sciences, or rather some natural extension of them, can in principle give a complete explanation for each and every bodily movement, or at least can do so up to whatever completeness is compatible with indeterminism in physics'' \cite[p.\,378]{jackson1996mental}.

These statements exemplify that the causal closure of the physical is generally taken to imply that every physical event which is explained at all, is explainable by natural science. But explanation, precisely construed~\cite{strevens2006scientific}, is only one way in which a theory can address events. Making room for prediction and other possible ways as well, we may take the above to imply that every physical event which is predicted, explained, or determined at all, can be predicted, explained, or determined by natural science. 

Applied to a theory of consciousness, this means that any physical event that the theory explains, predicts, or determines can (eventually) be explained, predicted, or determined by natural science. But for this to be true, the theory must not replace, alter, or add to the natural science account of physical events, because otherwise it would be committing itself to physical events that cannot be explained, predicted, or determined by natural science. Thus, the causal closure of the physical implies that a theory of consciousness cannot make changes to the physical events that are explained, predicted, or determined by natural science.

This point can be stated more clearly in formal terms. We have denoted the set of physical events that a theory of consciousness is committed to by $\P_T$. These are the events explained, predicted, or otherwise determined by that theory. And we have denoted the set of physical events explained, predicted, or otherwise determined by natural science (now or in the future) by $\P_P$. Thus, if every physical event that can be explained, predicted, or determined at all can be explained, predicted, or determined by natural science, then every event that is in $\P_T$ is also in $\P_P$. Taking into account the different languages that can be used in the two cases, this means that for every event description in $\P_T$ there is a corresponding event description of the same event in $\P_P$. This constitutes an injective function that maps $\P_T$ to $\P_P$.

We thus arrive at exactly the same formal requirement as in Equation~\eqref{eq:iota}. The causal closure of the physical implies that there is an embedding $\iota: \P_T \rightarrow \P_P$ that specifies for each physical event and physical description that the theory of consciousness is bound to the corresponding event and description explained, predicted, or determined by natural science.%
\footnote{
More advanced formulations of the causal closure of the physical lead to the same conclusion. Consider for example, the proposal by Barbara Montero and David Papineau in~\cite{montero2005defence}, that ``[e]very physical event is determined, in so far as it is determined
at all, by preceding physical conditions and laws''. Every physical event that is determined by preceding physical conditions and laws is an element of the class~$\P_P$. Every element of~$\P_T$ is, according to the broad reading of `determined' applied \cite{montero2005defence}, determined by a theory of consciousness. Hence it follows that every event in~$\P_T$ is also in~$\P_P$, and taking into account the different languages that may be used to describe the event, that there is an embedding~$\iota$ as in Equation~\eqref{eq:iota}.
}
Causal closure of the physical implies closure of the physical, and as a corollary of Theorem~\ref{thm:closure-unscientific} we posit that causal closure of the physical is also unscientific.%
\footnote{We note that the commonly understood epistemic reading of the closure of the physical, as expressed in Kim and Jackson's remarks, follows from the causal closure of the physical, as defined in the beginning of this section, only if an appropriate notion of `physical' is presupposed. This means that the causal closure of the physical must forbid the introduction of new physical entities that have effects that are not 
not explained by the physical sciences.
}

We emphasize that nowhere in our argument do we restrict to physical events which are already explained or predicted by natural science. 
What matters is only which relation a theory of consciousness proposes between the physical events it is committed to and the physical events that natural science posits.  Even if a theory presupposes that the physical events it associates with conscious experiences are determined by physical laws, but cannot in practice be explained or predicted based on these laws, as weak emergentist theories would have it, our argument applies. Theories of this sort may be wrong about what they say about physical events, and experiments may help to determine whether this is the case, but insofar as they buy into the very same underlying account of physical events as all other theories, the measurement results necessarily are the same as if any other theory were true. Because of the weak emergence claim, no postulate of such theory can imply any changes in the underlying physical events, and \emph{ipso facto} no changes in measurement results.%
\footnote{That is not so for strong emergentist theories, of course. These introduce genuine new causes and effects which are not claimed to be reducable to fundamental physical laws. It is well known that strong emergentist theories are not compatible with physicalism and the causal closure of the physical~\cite{sep-properties-emergent}.}

\section{Conclusion}\label{sec:final}

We have shown that the causal closure of the physical goes far beyond what is usually considered. Since all measurement results in the scientific study of consciousness are either physical events (such as keystrokes or sound waves) or at least determined by physical events (such as data stored on hard disks), no two theories obeying the causal closure of the physical can actually be distinguished in experiments. Our result applies to all major neuroscientific theories of consciousness as well as to the leading philosophical paradigms in the field. It applies to any theory of consciousness that fits into the natural science account of physical events without altering it. This includes all functionalist and identity theories of consciousness, such as GNW~\cite{mashour2020conscious}, HOT~\cite{brown2019understanding}, AST~\cite{graziano2015attention}, or predictive processing-based theories~\cite{schlicht2021you}, as well as eliminativist or illusionist theories~\cite{frankish2016illusionism}. But it also includes theories such as IIT, whose mathematics takes the form of a function that maps physical states and events to conscious states and events~\cite{kleiner2021mathematical}.%
\footnote{
Our results do not, however, apply to theories of temperature, life, or similar. They are fully compatible with there not being difficulties of the sort we point out in distinguishing different such theories empirically. 
Consider, as an example, the case of temperature, whose relation to microphysical events is sometimes claimed analogous to consciousness' relation to physics. In contrast to consciousness, however, experiments on temperature explore a purely macroscopic theory -- thermodynamics -- which does not address microphysics at all. The relation between temperature and microphysics is addressed only in terms of theory reduction of thermodynamics to statisctical physics~\cite{dizadji2010s}.
What is more, in statistical physics, the microphysical state actually depends on temperature, as apparent for example from the fact that temperature is part of the partition function that describes the state's statistical properties~\cite{landau1980course}.
If one were to change one's theory of how temperature supervenes on the physical, one would have to change these statistical properties as well so as to ensure the link to thermodynamics remains valid. Different theories of temperature are not compatible with one and the same microphysical distribution.
}

We have shown that no experiment of any kind can actually distinguish between these theories. Whatever measurement result is consistent with one theory is necessarily consistent with the other, because qua closure of the physical, the physical functioning of the brain, from stimulus presentation to verbal message or similar output, is exactly the same according to all these theories. This observation is at odds with the numerous experiments conducted to date to distinguish precisely between some of these theories. Our results show that there is a major flaw which underlies these experiments. The theories on which these experiments are based violate a necessary condition for the experiments to work as intended.

There are two potential conclusions that one can draw from our results. Either, experimenters do not really adhere to the closure of the physical when conducting experiments, but implicitly assume that the theories tested modify what falls solely within the realm of natural science. If this is the case, then our results constitute an imperative to improve the tested theories and make explicit what is implicitly assumed. If, on the other hand, experimenters do not implicitly adhere to the closure of the physical when running experiments, then our results call into question the very conclusions drawn on the basis of these experimental results. In either case, our results show that the closure of the physical must be abandoned in both theory and experiment. Theories of consciousness must explicitly state how what they take to be consciousness (physical or otherwise) comes to determine reports and other measures of consciousness, and to do this they must enter the realm of natural science.

In a very different context, Einstein once asserted that ``[it] is the theory which decides what we can observe'' ~\cite{filk2016theory,heisenberg1971physics}. It seems that this point has not yet been fully recognized in the construction of scientific theories of consciousness.

\subsection*{Acknowledgments} This research was supported by grant number FQXi-RFP-CPW-2018 from the Foundational Questions Institute and Fetzer Franklin Fund, a donor advised fund of the Silicon Valley Community Foundation. We would like to thank Sander Beckers, Ophelia Deroy, Alexander Gebharter, Kobi Kremnitzer, Christian List, Paul Taylor and Wanja Wiese for helpful discussions, and  David Chalmers, Joe Dewhurst, Timo Freiesleben, George Musser and Naftali Weinberger for feedback on an earlier draft. 


\end{document}